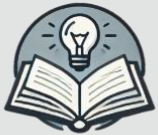

**THE PATENTIST**
LIVING LITERATURE REVIEW

---

# What is patent quality?

Gaétan de Rassenfosse

Holder of the Chair of Science, Technology, and Innovation Policy
École polytechnique fédérale de Lausanne, Switzerland.

---

This version: April 2025

**Purpose**
This article is part of a Living Literature Review exploring topics related to intellectual property, focusing on insights from the economic literature. Our aim is to provide a clear and non-technical introduction to patent rights, making them accessible to graduate students, legal scholars and practitioners, policymakers, and anyone curious about the subject.

**Funding**
This project is made possible through a Living Literature Review grant generously provided by Open Philanthropy. Open Philanthropy does not exert editorial control over this work, and the views expressed here do not necessarily reflect those of Open Philanthropy.

## What is patent quality?

Gaétan de Rassenfosse
École polytechnique fédérale de Lausanne, Switzerland.

Patent quality is a central concern for all stakeholders in the patent system. Inventors, patent attorneys, scholars, and policymakers consistently emphasize its importance, recognizing that the value and effectiveness of the patent system hinge on the issuance of high-quality patents.

But what exactly is patent quality? As we will see, the term is used to describe a wide range of distinct constructs. Conversely, the same construct may be described using different terminology, such as strength, importance, or value, adding another layer of confusion. This article explores the many facets of patent quality.

### Clarifying the foundation: What is invention quality?

Since patents are meant to protect inventions, defining what constitutes invention quality seems like a natural starting point. In the economics literature, invention quality is a stylized and abstract concept that captures 'how much better' a product or technology is compared to previous versions. It refers to the productivity-enhancing or utility-enhancing characteristics of that product or technology (*e.g.*, Grossman and Helpman 1991, Aghion and Howitt 1992).

While it is intuitive to conceive invention quality as a measure of an invention's technological superiority, this definition is not satisfactory for practical purposes: the notion of technical superiority is just as vague as the concept of quality itself. An alternative perspective emerges from the innovation literature, which distinguishes between types of technological advances, namely incremental vs. radical inventions. The former refers to improvements or refinements in known technologies, whereas the latter represents significant breakthroughs with strong disruptive potential, such as the invention of the first 3D printer. In addition, a restricted number of inventions have a widespread, transformative impact on the entire economy, often leading to long-term changes in how we live and work—for instance, the steam engine, electricity, and the internet. We call these inventions general-purpose technologies (*e.g.*, Bresnahan and Trajtenberg 1995).

While this perspective offers a gradation of inventions' importance, it does not provide a concrete way to assess quality improvements across successive generations of inventions. Such an assessment requires acknowledging the multidimensional and context-dependent nature of quality.

Consider the case of a piezoelectric sensor, which converts mechanical pressure into an electrical signal. This single invention can be embedded in a wide array of products, each placing different demands on its performance. In electronic drum pads, for instance, the sensor is used to detect strikes and trigger digital sounds. Relevant dimensions of quality include responsiveness (how accurately the sensor captures input), latency (how quickly the signal is processed), and durability (its ability to withstand repeated physical stress). In contrast, when embedded in a car's airbag system, the same sensor must detect sudden shocks to trigger deployment. Here, while responsiveness and latency still matter, other dimensions—such as resistance to temperature fluctuations and humidity—become equally critical. Piezoelectric sensors are also found in a variety of everyday products, including electric lighters, digital bathroom scales, and inkjet printers. In each case, quality is assessed over multiple dimensions, and improvements along one dimension may be essential in one application but irrelevant in another.

This example illustrates the challenge of measuring invention quality with a single, unified metric. Yet, at least in principle, it is possible to assess an invention's quality ex-ante by identifying its relevant technical dimensions—such as efficiency, durability, or precision—and evaluating them under controlled conditions, even in the absence of a concrete use case. (This is particularly true for incremental inventions, as radical inventions may introduce entirely new quality dimensions, leaving no baseline for measuring improvements over previous generations of the technology.)

**What about invention value?**

While invention quality can be assessed in the absence of concrete use, invention value is inherently tied to use. An invention has economic value only insofar as it is, or can be, embedded in a product or process that generates impact in the real world. Unlike quality, however, value lends itself—at least theoretically—to a single, comparable measure: its monetary worth. While measuring quality requires navigating multiple, often incommensurable dimensions, value can be expressed in real money.

An important corollary to this statement is that different uses for an invention will lead to different valuations. Consider the case of the ring-pull can. The inventor allegedly [licensed the system to Coca-Cola](#) at 0.1 pence per can and obtained £148,000 a day in royalties. Had the inventor granted an exclusive license for his invention to a company specializing only in canned green beans, he would have amassed far less money—highlighting that the actual use of an invention directly affects its value.

This example focuses on what economists call the 'private' value of the invention, namely, the returns accruing to the innovator. The private value forms only one component of an invention's social value, which also encompasses the [consumer surplus](#) (capturing the fact that some inventors may have paid less than what they were willing to pay), [spillovers](#) to competitors and other actors such as suppliers (who may capture some of the value created by the invention), and wider societal effects (such as environmental benefits).

While the private value of an invention is tied to how a particular firm chooses to use it, the social value of an invention reflects its broader potential—across users, applications, and contexts. The inventor of the ring-pull can may have initially licensed it for use in canned beverages, but as other companies adopt the invention for canned vegetables, soups, or even pet food, its benefits diffuse throughout the economy. Consumers gain from improved convenience, competitors are nudged to innovate, and suppliers may benefit from increased demand or efficiency gains. These spillovers, along with broader societal impacts—such as improvements in accessibility—contribute to the social value of the invention.

The distinction between private and social value highlights a central insight: while private value depends on specific market conditions and the innovator's approach to appropriate returns, social value emerges more diffusely as others adopt and adapt the invention. As a result, the realized private value of an invention may fall well short of its potential, whereas the realized social value may come closer to being fully realized—precisely because it is not restricted to a single use or actor.

Finally, note that the term 'value' is not confined to the 'economic merit' of the invention. A previous article introduced the notions of 'use value' and 'exchange value.'[1] Use value refers to an object's utility or practical usefulness, whereas exchange value refers to what it can be traded for in the market. The former concept more closely relates to the notion of invention quality, while the latter refers to the invention's economic value (capturing one dimension of the social value).

**So, where does patent quality fit in?**

We have taken a long detour in our effort to understand what makes a patent 'high quality.' A natural temptation is to equate patent quality with the quality of the underlying invention. While some scholarly works adopt this definition, patent quality can, in fact, be understood in several distinct ways, as discussed by Guerrini (2014), Schmitt (2025), and others. It is entirely possible for a high-quality patent to protect a low-quality invention, just as a low-quality patent may cover a groundbreaking one. There are at least four broad definitions of patent quality.

As just mentioned, the first definition conflates the quality of the patent with the *quality of the invention it protects*. In this context, invention quality is measured, for instance, by its novelty, inventiveness, or the size of the 'inventive step' separating the invention from existing technology. This abstract, standardized metric of quality is useful for stylized economic modeling but fails to recognize the multidimensional and context-dependent nature of quality, as the example of the piezoelectric sensor illustrates.

A second perspective links patent quality to the *economic value or impact of the patented invention*. From this standpoint, a good patent is one that fulfills the key objectives of the patent system: to reward and incentivize innovation while promoting knowledge diffusion and follow-on innovation. Accordingly, high-quality patents are those that generate significant social value.

A third definition focuses on the drafting quality of the patent. In this view, a high-quality patent is a *well-written document*, especially in terms of claim clarity and appropriate scope. It provides a clear and comprehensive description of the invention, and its claims are precise and broad enough to offer strong protection.

A fourth perspective defines quality in terms of *compliance with statutory requirements*. A high-quality patent is one that meets the criteria of patentability: subject matter eligibility, utility, novelty, non-obviousness, enablement, and written description. Here, a low-quality patent is one that should not have been granted in the first place. Conversely, a high-quality patent is one that is likely to be legally valid if challenged. The USPTO largely adopts this [legalistic definition](#), focusing its quality initiatives on identifying and minimizing the issuance of potentially invalid patents.

---

[1] de Rassenfosse, G. (2025). What proportion of knowledge is patented? The Patentist Living Literature Review 2: 1–4. DOI: 10.48550/arXiv.2501.18043.

It is worth emphasizing that no single, universally accepted definition of patent quality exists. Indeed, each profession may have its favorite definition: patent attorneys may prioritize clarity and breadth of claims, engineers may value technological sophistication, and economists may focus on the invention's impact.

**What about patent value?**

Patent value refers to a patent's potential to deliver economic returns or strategic benefits to its owner. Value is shaped by the existence and size of the relevant market—the actual use of inventions. Selling the ring-pull can patent to Coca-Cola, for example, clearly yields more value than licensing it to a niche soup manufacturer.

It would be erroneous to conflate patent value with the value of the underlying invention. Indeed, patent protection creates *additional* value due to the exclusionary rights over the invention. The mere existence of a patent can enhance the private value of an invention. Take the ring-pull can again: without a patent, selling the invention to Coca-Cola would yield limited benefits—rivals could freely copy the invention, eroding any first-mover advantage and driving the private value of the invention to zero. By contrast, the patent allows the would-be buyer to obtain exclusive rights, increasing the invention's marketability. This added value is known in the literature as the 'patent premium.' One estimate using data from a survey of Australian inventors puts it at 40 to 50 percent (Jensen et al. 2011).

As noted in a previous article, firms file patents for many reasons and employ them in varied ways—whether to block competitors, attract investment, or strengthen bargaining positions.[2] These different uses represent as many value-creation strategies that go beyond the invention's worth, further explaining why the value of the invention differs from the value of the associated patent right.

Crucially, while the existence of the patent right adds value, the actual value realized depends on the quality of the patent—particularly its legal strength. Because patentees typically need to prove infringement and withstand invalidity challenges in court, legal validity is a core component of patent value. In that sense, quality and value are tightly intertwined.

However, Farrell and Shapiro (2008) explain that patents with dubious validity ('weak' patents) can still command substantial royalty payments. Even if a downstream firm suspects that a patent is weak, the cost of proving its invalidity in court can be substantial, making a settlement with the patent holder (through royalty payments) a more economically rational option. Thus, even weak patents can shape—or distort—markets, making high-quality patent examination essential to a well-functioning innovation system.

---

[2] de Rassenfosse, G. (2025). Why do firms patent? The Patentist Living Literature Review 3: 1–5. DOI: 10.48550/arXiv.2503.11555.